\documentclass[useAMS,usenatbib]{mn2e} 

\usepackage{amssymb,amsmath}
\usepackage{psfig,epsfig,amssymb,alltt}
\usepackage{abbrevjournaux}
\usepackage{latexsym}
\psfigurepath{./PS:.}
\vspace{1cm}

\title[Cosmological constraints from the capture of non-Gaussianity in Weak Lensing data]{Cosmological constraints from the capture\\ of non-Gaussianity in Weak Lensing data}
\author[S.~Pires, A.~Leonard and J.-L.~Starck]
  {Sandrine Pires\thanks{Email: sandrine.pires@cea.fr}, Adrienne Leonard and Jean-Luc Starck\\
 Laboratoire AIM, CEA/DSM-CNRS-Universite Paris Diderot, IRFU/SEDI-SAP, Service d'Astrophysique, \\CEA Saclay, Orme des Merisiers, 91191 Gif-sur-Yvette, France}
  \date{Released 2011 Xxxxx XX}

\pagerange{\pageref{firstpage}--\pageref{lastpage}} \pubyear{20112}

\def\LaTeX{L\kern-.36em\raise.3ex\hbox{a}\kern-.15em
    T\kern-.1667em\lower.7ex\hbox{E}\kern-.125emX}

\begin{document}

\label{firstpage}

\maketitle

\begin{abstract}
Weak gravitational lensing has become a common tool to constrain the cosmological model.
The majority of the methods to derive constraints on cosmological parameters use second-order statistics of the cosmic shear. 
Despite their success, second-order statistics are not optimal and degeneracies between some parameters remain.
Tighter constraints can be obtained if second-order statistics are combined with a statistic that is efficient to capture non-Gaussianity.
In this paper, we search for such a statistical tool and we show that  there is additional information to be extracted from statistical analysis of the convergence maps beyond what can be obtained from statistical analysis of the shear field.
For this purpose, we have carried out a large number of cosmological simulations along the $\sigma_8$-$\Omega_m$ degeneracy,
and we have considered three different statistics commonly used for non-Gaussian features characterization: skewness, kurtosis and peak count.
To be able to investigate non-Gaussianity directly in the shear field we have used the aperture mass definition of these three statistics for different scales.
Then, the results have been compared with the results obtained with the same statistics estimated in the convergence maps at the same scales. 
First, we show that shear statistics give similar constraints to those given by convergence statistics, if the same scale is considered.
In addition, we find that the peak count statistic is the best to capture non-Gaussianities in the weak lensing field and to break the $\sigma_8$-$\Omega_m$ degeneracy. We show that this statistical analysis should be conducted in the convergence maps:
first, because there exist fast algorithms to compute the convergence map for different scales, and secondly because it offers the opportunity to denoise the reconstructed convergence map, which improves non-Gaussian features extraction.
\end{abstract}


\section{Introduction}
Gravitational light deflection, caused by large scale structure along the line-of-sight, produces an observable pattern of alignments in the images of distant galaxies.
This distortion of the images of distant galaxies by gravitational lensing, called cosmic shear, offers an opportunity to directly probe the total matter distribution of the Universe, and not just the luminous matter.
Therefore, the statistical properties of this gravitational shear field are directly linked to the statistical properties of the total matter distribution and can thus be directly compared to theoretical models of structure formation. 
Despite some systematics (PSF distortion, intrinsic alignments...), this approach is extremely attractive since it is unaffected by the biases characteristic of methods based only on the light distribution.

Since its first detection \citep{wlens:vanwaerbeke00,wlens:kaiser00,wlens:wittman00, wlens:bacon00}, cosmic shear has rapidly become a major tool to constrain the cosmological model \citep[for review, see e.g.][]{wlens:mellier99,wlens:bartelmann01,wlens:refregier03,wlens:hoekstra08,wlens:munshi08}.
 
In most weak lensing studies, second-order statistics are the most commonly used statistical probe \citep[e.g.][]{stat:maoli01,stat:hoekstra06,stat:benjamin07,map:fu08} because of their potential to constrain the power spectrum of density fluctuations in the late Universe.
However, second-order statistics are not optimal to constrain cosmological parameters. 
For example, they 
only depend on a degenerate combination of the amplitude of matter fluctuations $\sigma_8$ and the matter density parameter $\Omega_m$ \citep{stat:maoli01,stat:refregier02,stat:bacon03,stat:massey05,stat:dahle06}.

The optimality of second-order statistics to constrain cosmological parameters depends heavily on the assumption of Gaussianity of the field.
However the weak lensing field is composed, at small scales, of non-Gaussian features such as clusters of galaxies.
These non-Gaussian signatures, which can be measured via higher-order moments, carry additional information that cannot be extracted with second-order statistics. Since the non-Gaussianity is induced by the growth of structures, it holds important cosmological information.
Many studies \citep[e.g.][]{combine:bernardeau97,combine:takada04,map:kilbinger05,stat:pires09a,stat:berge10} have shown that combining second-order statistics with higher-order statistics tighten the constraints on cosmological parameters.

Most non-Gaussian studies \citep[e.g.][]{map:schneider98,map:jarvis04,map:kilbinger05,map:dietrich10} have been performed in the shear field because it can be directly derived from the shape of galaxies.
This paper aims to produce evidence that there is additional information to be extracted from a higher-order statistical analysis of the convergence maps 
beyond what can be obtained from a higher-order statistical analysis of the shear field because higher-order statistics are probing the non-Gaussian features of the signal and these non-Gaussian structures can be  better reconstructed in convergence maps using a denoising. 
In \cite{stat:pires09a}, the efficiency of several higher-order convergence statistics have been compared to discriminate cosmological models along the $\sigma_8$-$\Omega_m$ degeneracy. In this paper, we are interested in showing the advantage of using these higher-order convergence statistics compared to higher-order shear statistics. 
This comparison cannot be performed directly because the evaluation of non-Gaussian statistics in the shear field requires to use their aperture mass definition \citep{map:schneider98} or another different filter that is defined for a given scale $\theta$. A fair comparison with convergence statistics requires the statistics in the convergence maps to be estimated at the same scale.
A stationary wavelet transform, the "\`a trous" wavelet transform, has been used in this paper to compute the convergence statistics at the given scale $\theta$.


The paper is organized as follows. In section 2, the cosmological models selected for this study are described, followed by a short description of the weak lensing simulations. Section 3 summarizes the different statistics used in this study. We give the definition of the aperture mass $M_{ap}$ and present the three shear statistics considered in this study. Then, the "\`a trous" wavelet transform is defined, as well as the three related convergence statistics.
Section 4 presents our results and we summarize our conclusions in section 5. 

\section{Simulations of weak lensing mass maps}
\label{sect_sim}

\begin{figure}
\centerline{
\includegraphics[height=6.cm,width=6.cm]{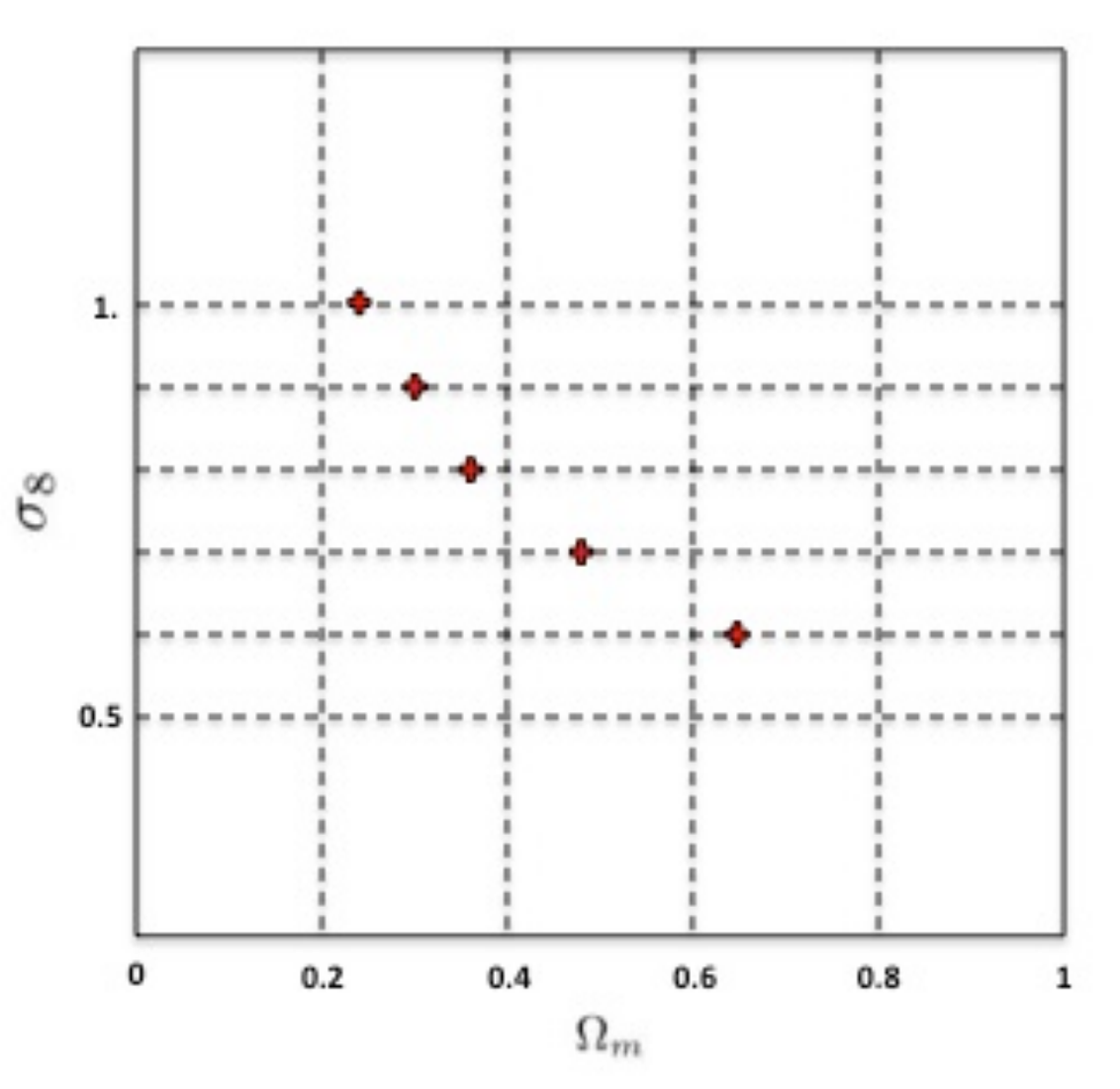}
}
\caption{Location of the 5 simulated cosmological models in the $\sigma_8$-$\Omega_m$ plane. 
The 5 cosmological models have been selected along the $\sigma_8$-$\Omega_m$ degeneracy corresponding to current constraints from a power spectrum analysis.}
\label{model}
\end{figure}

N-body simulations have been used to numerically compute the variation of the different statistics with cosmological parameters
and then compare their ability to break degeneracies. We carried out N-body simulations for 5 different cosmological models along the $\sigma_8$-$\Omega_m$ degeneracy corresponding to current constraints from a power spectrum analysis. The N-body simulations were carried out using the RAMSES code \citep{code:Teyssier02}.
Fig. \ref{model} shows the distribution of these cosmological simulations in the $\sigma_8$-$\Omega_m$ plane. The characteristics of these cosmological models have been given in \cite{stat:pires09a}. For each cosmological model, we have run 100 realizations in order to quantify the observational uncertainties.\\

In the N-body simulations that are commonly used in cosmology, the dark matter distribution is represented by discrete massive particles. The simplest way of treating these particles is to map their positions onto a pixelized grid. In the case of multiple sheet weak lensing, we do this by taking slices through the 3D simulations. These slices are then projected into 2D mass sheets.  
The effective convergence can subsequently be calculated by stacking a set of these 2D mass sheets along the line of sight, using the lensing efficiency function. This is a procedure that was used before by \cite{code:vale03}, where the effective 2D mass distribution $\kappa_e$ is calculated by integrating the density fluctuation along the line of sight. 
Using the Born approximation, which assumes that the light rays follow straight lines, the convergence can be numerically expressed by
\begin{equation}
\label{kappae}
\kappa_e \approx \frac{3H_0^2\Omega_m L}{2c^2}\sum_i\frac{\chi_i(\chi_0-\chi_i)}{\chi_0a(\chi_i)}\left(\frac{n_pR^2}{N_ts^2}-\Delta r_{f_i} \right),
\end{equation}
where $H_0$ is the Hubble constant, $\Omega_m$ is the density of matter, $c$ is the speed of light, $L$ is the length of the box, and $\chi$ are comoving distances where $\chi_0$ is the   distance to the source galaxies. The summation is performed over the $i^{th}$ box. The number of particles associated with a pixel of the simulation is $n_p$, the total number of particles within a simulation is $N_t$, and $s=L_p/L$, where $L_p$ is the length of the plane responsible for the lensing. $R$ is the size of the 2D maps and $\Delta r_{f_i} = \frac{r_2-r_1}{L}$, where $r_1$ and $r_2$ are comoving distances.

\begin{figure}
\centerline{
\includegraphics[height=7.cm,width=6.cm]{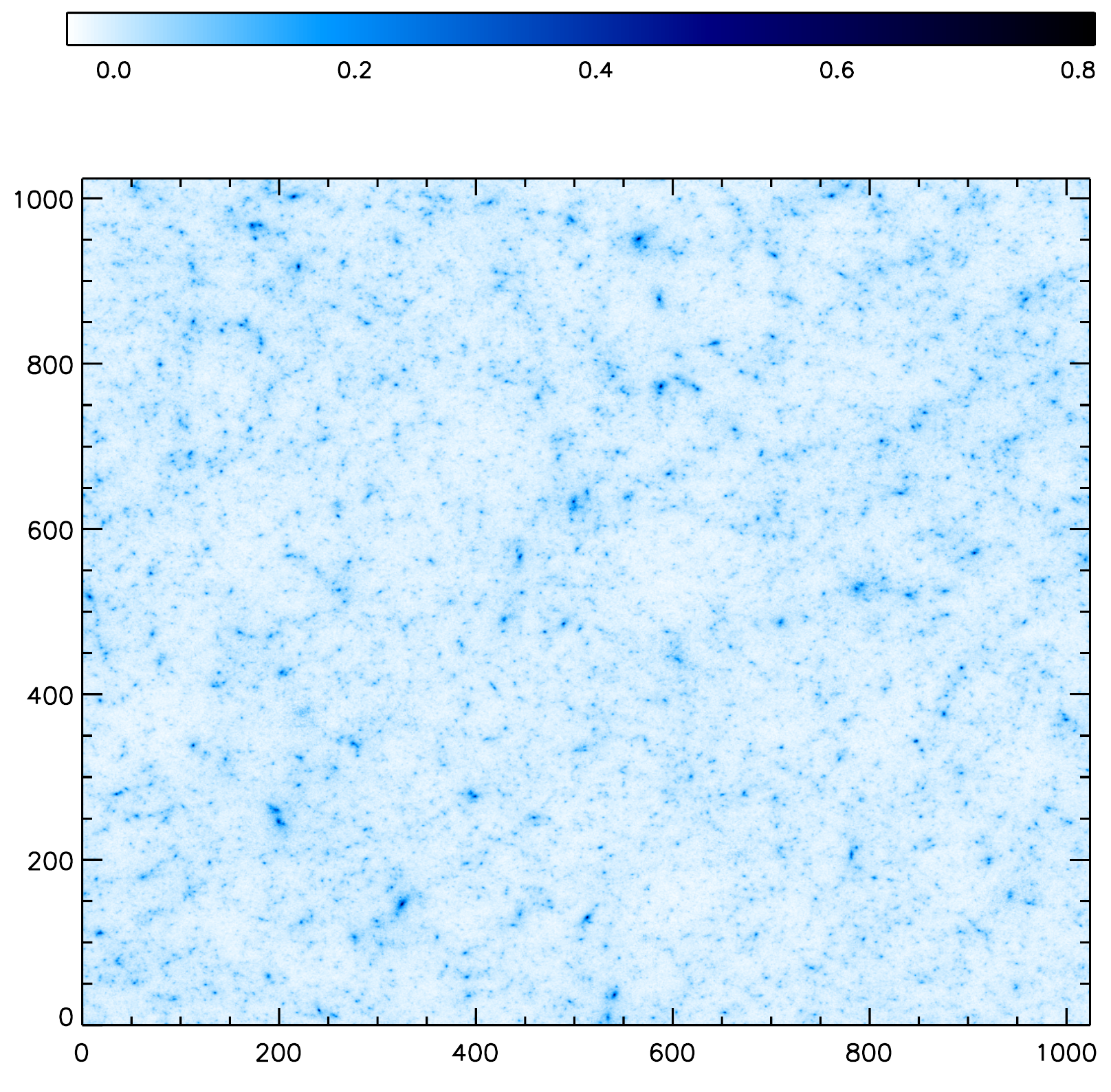}
}
\caption{Simulated convergence map corresponding to a realization of a cosmological model with parameters: $\Omega_m = 0.23$, $\Omega_L = 0.77$, $h = 0.594$, $\sigma_8 = 1$. The field is $3.95^{\circ}$ x $3.95^{\circ}$ downsampled to $1024$ x $1024$ (pixel scale = $0.23\arcmin$).}
\label{kappa}
\end{figure}

For each of our 5 models, we have run 21 N-body simulations, each containing $256^3$ particles. 
We have used these 3D N-body simulations to derive 100 realizations of the convergence field for each model.
The field is $3.95^{\circ}$x $3.95^{\circ}$ and it has been downsampled to $1024$ x $1024$ pixels (1 pixel = 0.23$\arcmin$). Figure \ref{kappa} shows one of these convergence maps. As said previously, the 5 cosmological models have been selected along the $\sigma_8$-$\Omega_m$ degeneracy corresponding to current constraints from a power spectrum analysis. Additionally, we have verified that the power spectrum of the 5 cosmological models are still degenerated with the survey area considered.

The convergence map $\kappa$ that corresponds to the projected matter density is not directly observable 
but, it can be derived from the observed shear maps $\gamma$ using the following relation \citep{wlens:kaiser93,wlens:starck06}:
\begin{eqnarray}
\label{eqn_reckE}
&&\hat{\kappa}  =  \hat{P_1} \hat{\gamma}_{1}+ \hat{P_2}\hat{\gamma}_{2}, \\
\textrm{with}:\nonumber\\ 
&&\hat{P_1}({\bf k}) = \frac{k_1^2-k_2^2}{k^2}\nonumber,\\
&&\hat{P_2}({\bf k}) = \frac{2 k_1 k_2}{k^2}\nonumber,
\end{eqnarray}
where the hat symbol denotes Fourier transform.
Inversely, the shear maps $\gamma_i$ can easily be derived from the convergence map using the following relation:
\begin{equation}
\label{eqn_gamma}
\hat{\gamma}_i = \hat{P}_i \hat{\kappa}
\end{equation}

In practice, observed shear maps $\gamma_i^{obs}$ are obtained by averaging over a finite number of galaxies and are therefore noisy: $\gamma_{i}^{obs} = \gamma_i + N_i$, where $N_1$ and $N_2$ are noise contributions with zero mean and standard deviation $\sigma_n=\sigma_{\epsilon}/ \sqrt{A.n_g}$ with $A$ being the area of the pixel in arcmin$^2$ and $n_g$ the average number of galaxies per arcmin$^2$. Typical values for the density of galaxies is $n_g = 30$ gal/arcmin$^2$ for ground-based simulations and  $n_g = 100$ gal/arcmin$^2$ for space-based simulations.
Although a bit optimistic, these two configurations have been considered to derive noisy shear maps and to compute the shear statistics described in \S \ref{sect_shearstat}. The derived noisy shear maps are downsampled to $1024$ x $1024$ pixels  (1 pixel = 0.23$\arcmin$) like the simulated convergence maps.
An estimation of the corresponding noisy convergence maps can be derived from the equation [\ref{eqn_reckE}]:
$ \hat{\kappa}_n= \hat{\kappa} + \hat{N}$,
where $\hat{N} = \hat{P_1} \hat{N_1} + \hat{P_2} \hat{N_2}$.
As follows, the noise $N$ in $\kappa_n$ is still Gaussian and uncorrelated.  
The inversion does not amplify the noise, but $\kappa_n$ may be dominated by the noise if $N$ is large, which is the case in practice.
Ground-based and space-based simulations of convergence maps have been derived by this way. The noisy convergence maps  derived by inversion are still $3.95^{\circ}$x $3.95^{\circ}$ downsampled to $1024$ x $1024$ pixels  (1 pixel = 0.23$\arcmin$) and they have been used to compute the convergence statistics described in \S  \ref{sect_mrlensstat}. \\

\section{Weak Lensing statistics}
\label{sect_stat}
The properties of the shear field $\gamma$ (or associated convergence field $\kappa$) can be measured statistically and reveal precious information about cosmological parameters. 
Up to now, cosmic shear studies have focused mainly on second-order statistics which only probe the Gaussian part of the matter distribution. 
However, the matter distribution is composed of non-Gaussian features such as the clusters of galaxies that are the result of the non-linear evolution of the primordial Gaussian field.
Therefore, higher-order statistics are required to probe the non-Gaussian part of the field and thus improve our constraints on cosmological parameters. 
This study will focus on these higher-order statistics estimated both in the shear maps $\gamma$ and in the convergence maps $\kappa$.



\subsection{Shear statistics}
\label{sect_shearstat}

The shear field $\gamma$ can be directly derived from measurements of the shape of galaxies. 
For this reason, two-point statistics of the shear field have become the standard way of constraining cosmological parameters \citep[see for example ][]{stat:maoli01,stat:hoekstra06,stat:benjamin07,map:fu08}.
Most of the interest in this type of analysis comes from its potential to constrain the spectrum of density fluctuations present in the late Universe and thus the cosmological parameters.
However, as said previously, cosmological parameters cannot be determined accurately using only second-order statistics because only the Gaussian features of the field are captured by this method.
Therefore, higher-order statistics have been introduced to probe the non-Gaussian features of the field and thus break degeneracies.
However, although the two-point correlations of the spin-2 shear field $\gamma_i$ can be reduced to a scalar quantity for parity reasons, 
this is not the case for higher-order moments of the shear field \citep[see][]{map:schneider03,map:takada03,map:zaldarriaga03}.
A way to get round this problem is to estimate higher-order statistics of the aperture mass $M_{ap}$, which has been introduced by \cite{map:schneider96}, rather than using the shear field directly.\\


The {\bf aperture mass $M_{ap}$} is one of the most widely used techniques for probing non-Gaussianity from the shear field \citep[e.g.][]{map:schneider98,map:jarvis04,map:kilbinger05,map:dietrich10}. 

The aperture mass $M_{ap}$ can be expressed in terms of the tangential component of the shear $\gamma_t$:
\begin{equation}
M_{ap}(\theta) = \int d^2 {\bf \vartheta} Q_{\theta} (\vartheta) \gamma_t({\bf \vartheta}),
\label{map_shear}
\end{equation}
where $Q_{\theta}(\vartheta)$ is a radially symmetric, finite and continuous weight function and ${\bf \vartheta}$ is measured from the center of the aperture. The choice of the weight function $Q_{\theta}(\vartheta)$ is arbitrary at this point.
In this paper, we have used the form introduced by Crittenden et al. 2002, that has been found to be more sensitive for constraining $\Omega_m$ than other forms (Zhang et al, 2003) :
\begin{equation}
Q_{\theta}(\vartheta) = \frac{\vartheta^2}{4 \pi \theta^4} \exp\left( -\frac{\vartheta^2}{2 \theta^2}\right).
\label{aperture_shear}
\end{equation}
In contrast with the shear field, which is a spin-2 field from which higher-order moments are not trivial to define, 
the  aperture mass defined by equation [\ref{map_shear}] is a scalar (spin-0) field, the skewness and kurtosis of which are well defined.\\

\begin{figure*}
\vbox{
\centerline{
\hbox{
\includegraphics[height=7.cm,width=6.cm]{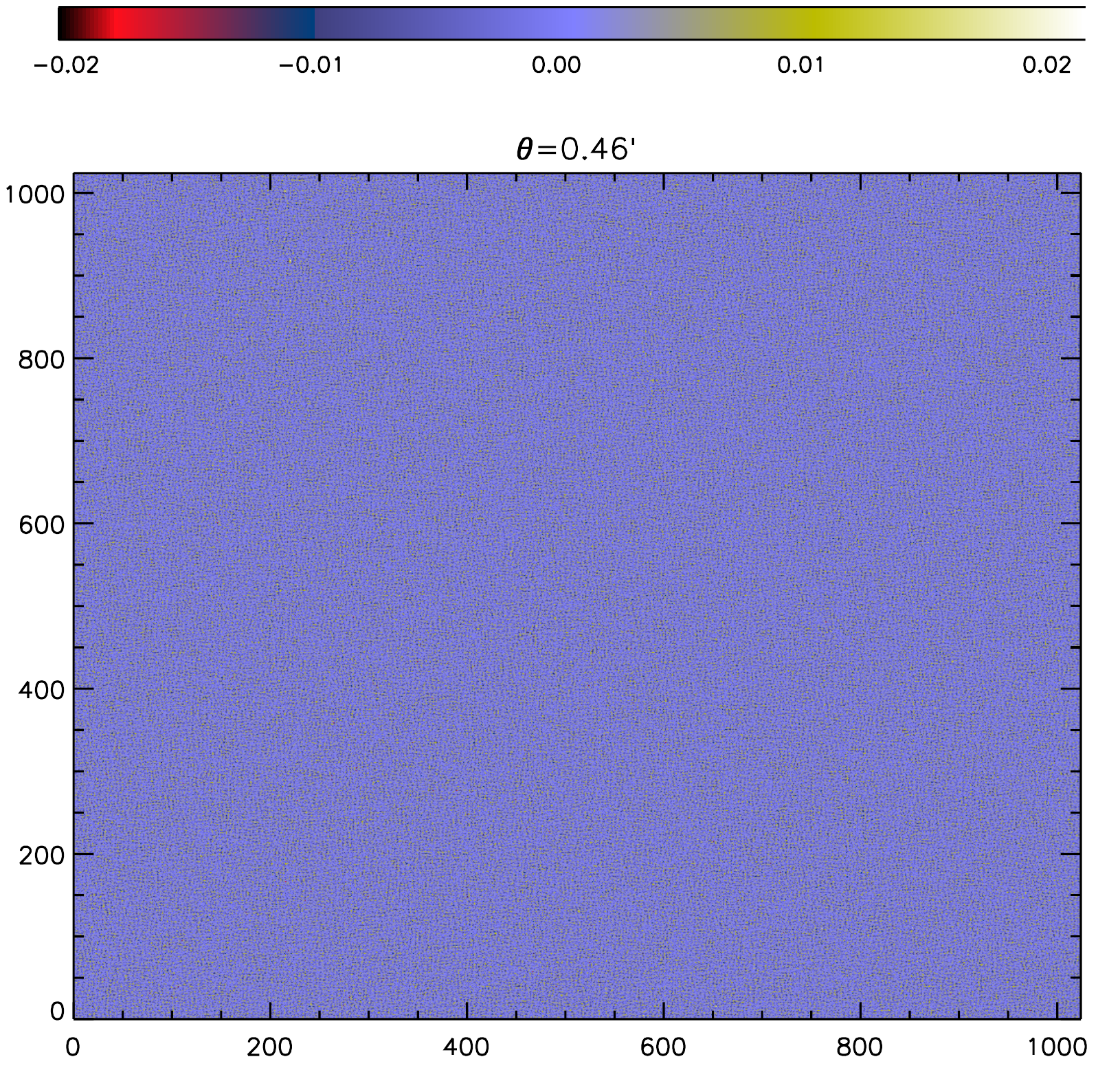}
\hspace{0.2cm}
\includegraphics[height=7.cm,width=6.cm]{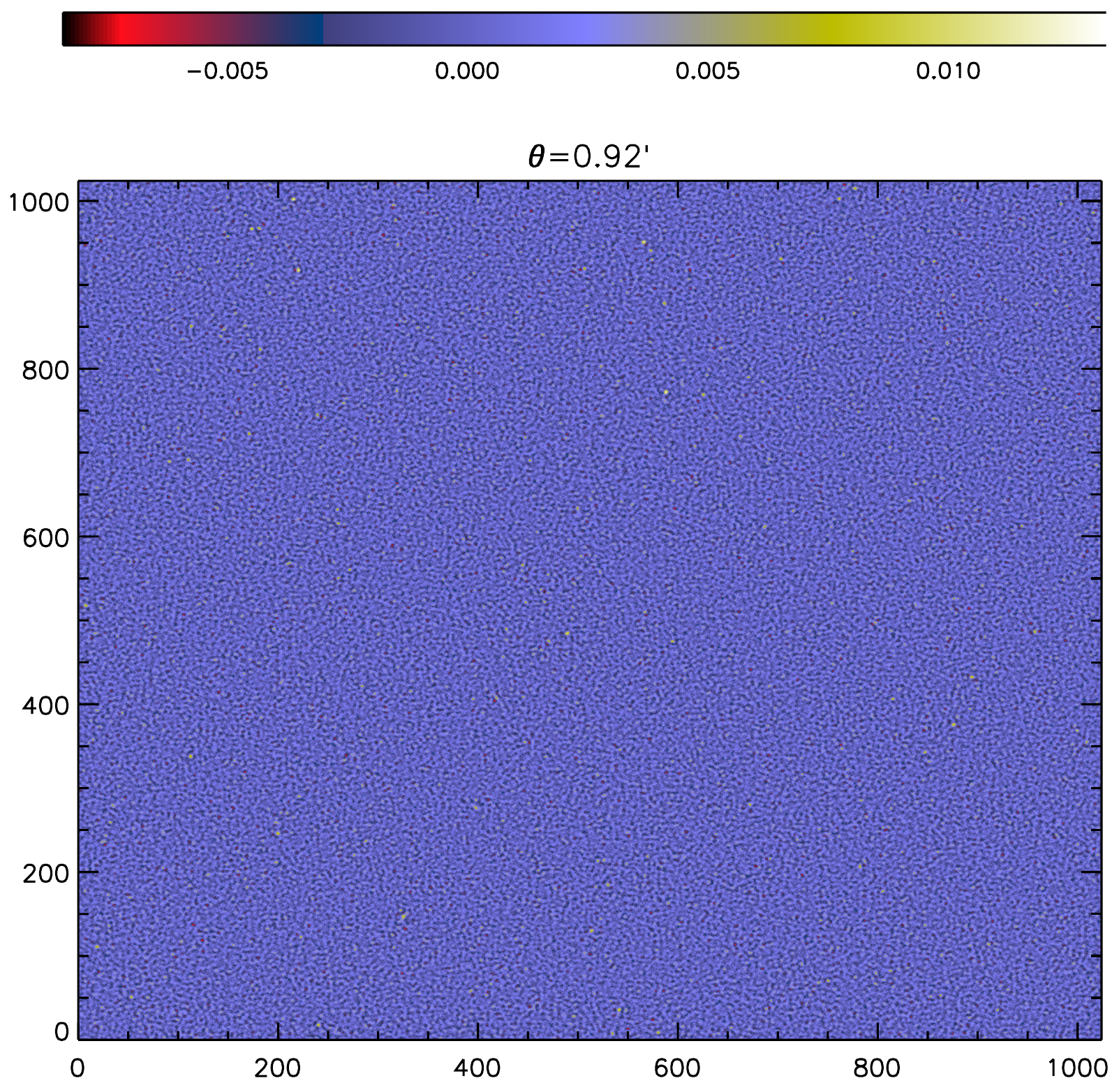}
}}
\vspace{0.3cm}
\centerline{
\hbox{
\includegraphics[height=7.cm,width=6.cm]{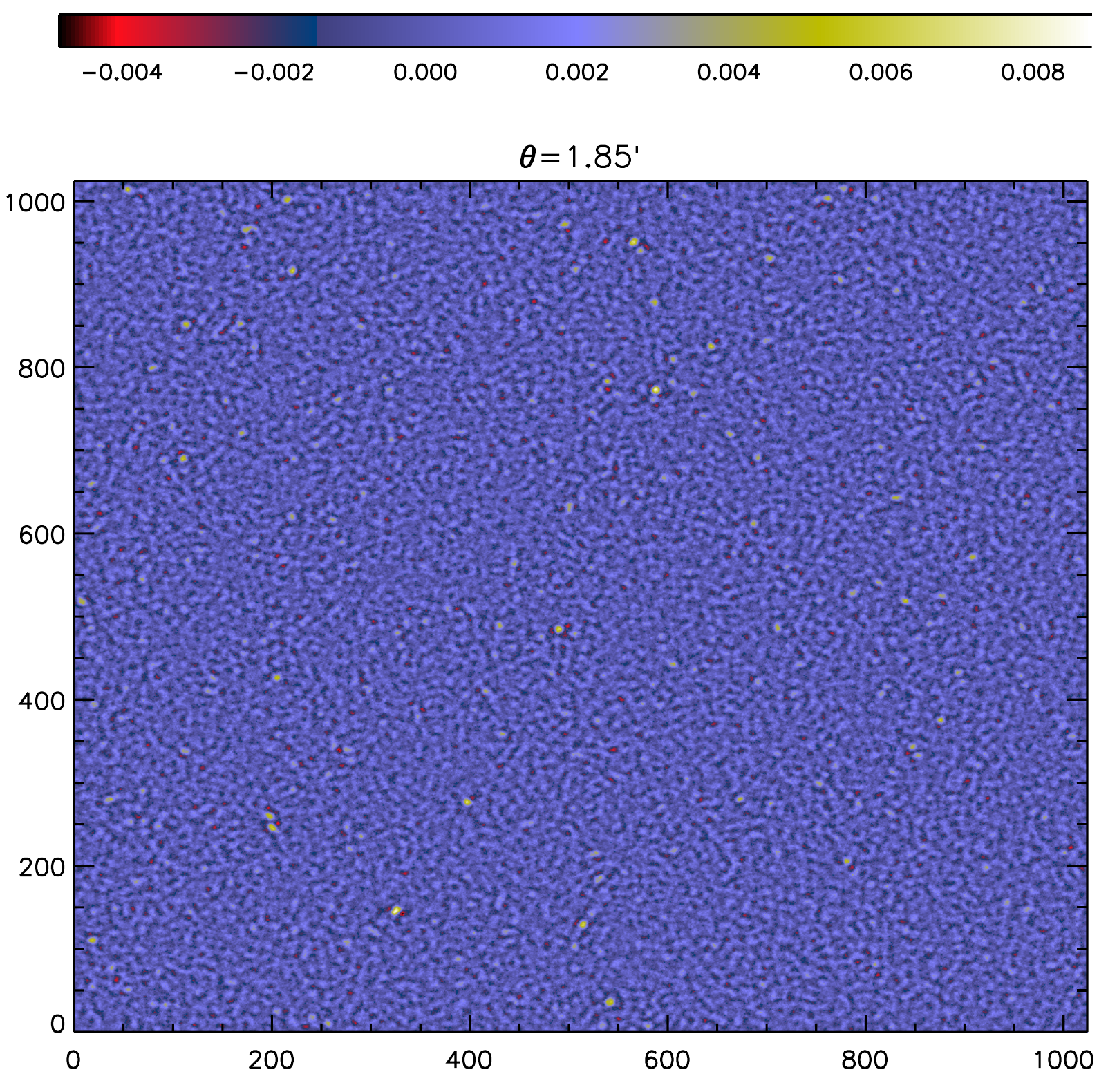}
\hspace{0.2cm}
\includegraphics[height=7.cm,width=6.cm]{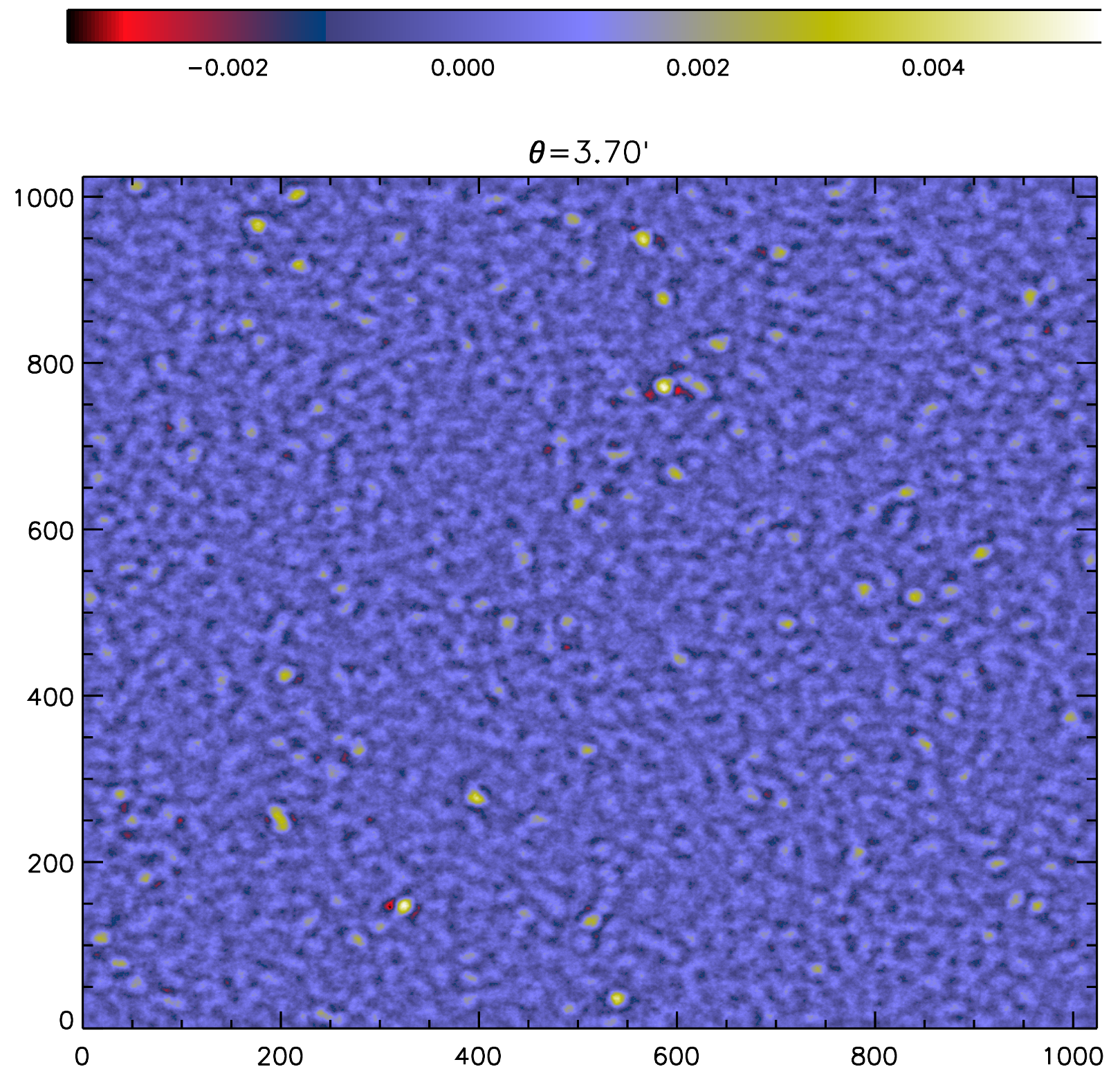}
}}
\vspace{0.3cm}
\centerline{
\hbox{
\includegraphics[height=7.cm,width=6.cm]{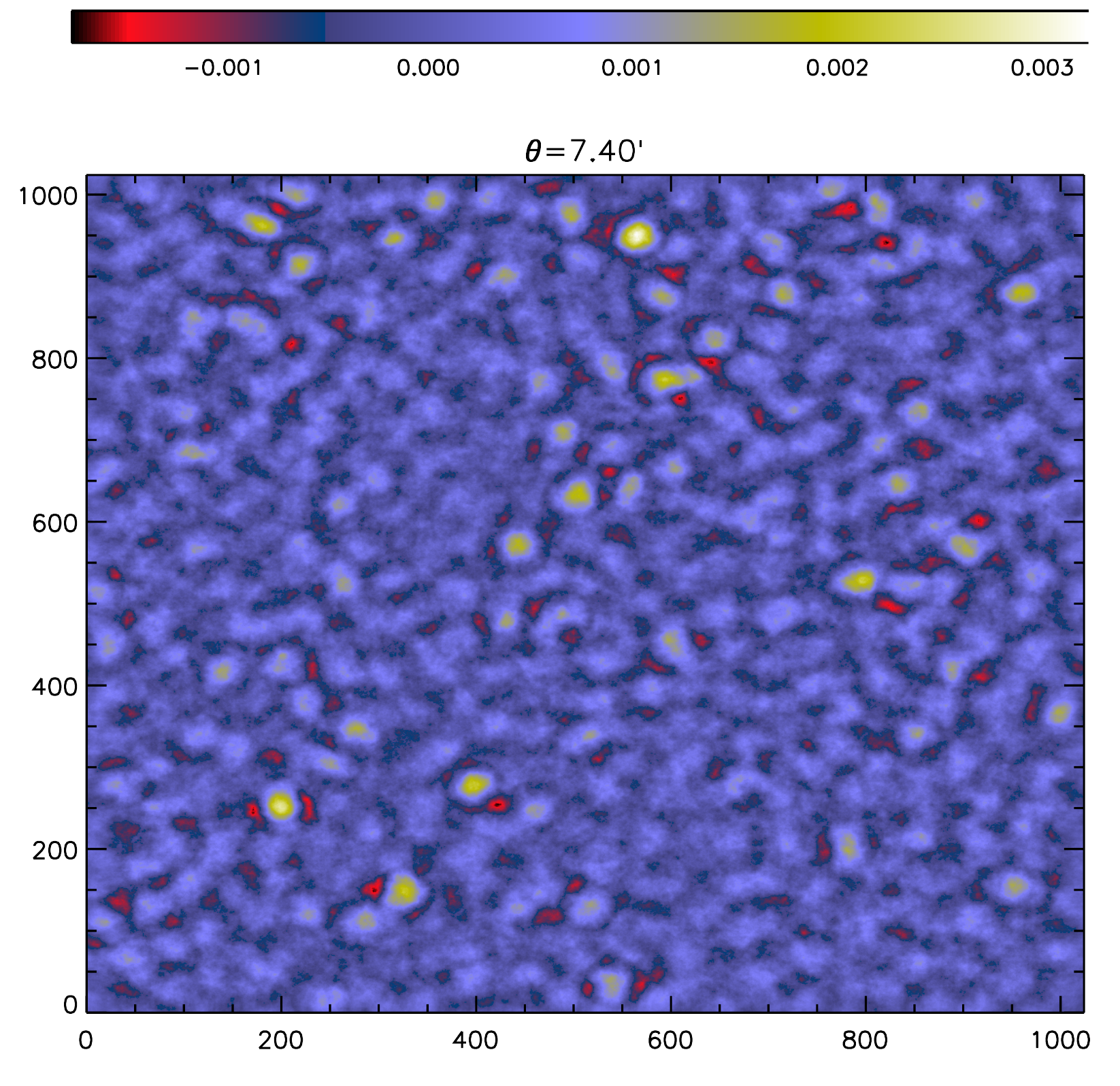}
\hspace{0.2cm}
\includegraphics[height=7.cm,width=6.cm]{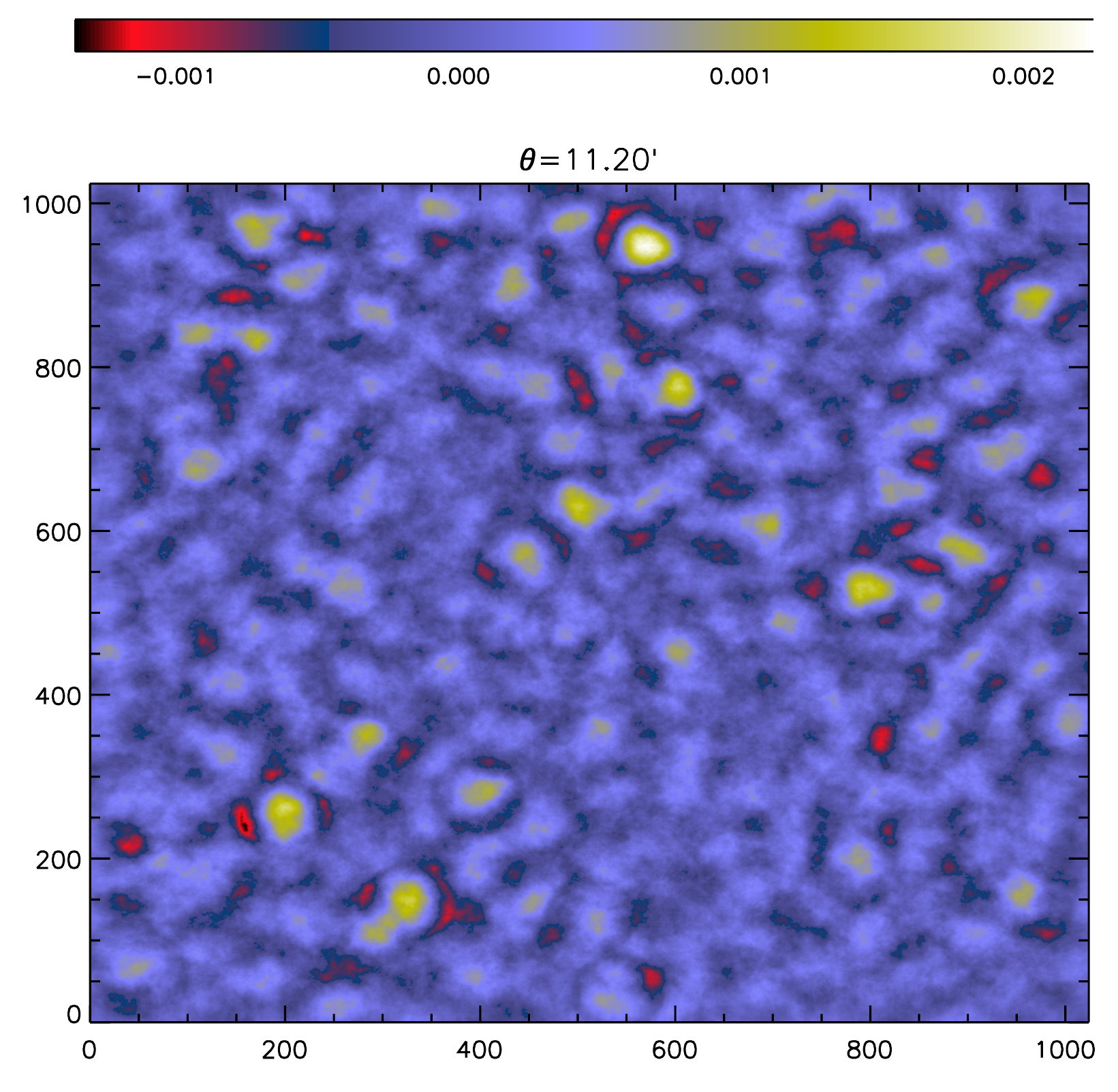}
}}
}
\caption{Aperture mass maps obtained from noisy shear maps (space-based simulations) filtered with an aperture mass with scales $\theta = 0.46', 0.92', 1.85', 3.70', 7.40', 11.20'$. The field is $3.95^{\circ}$ x $3.95^{\circ}$ downsampled to $1024$ x $1024$ i.e. a pixel scale of $0.23\arcmin$.}

\label{map}
\end{figure*}

In this section, for each noisy shear maps $\gamma$ described in \S \ref{sect_sim}, we have estimated the aperture mass $M_{ap}(\theta)$ for several apertures $\theta$ from the relations [\ref{map_shear}] and  [\ref{aperture_shear}]. Figure \ref{map} shows some of these aperture mass maps for apertures $\theta = 0.46', 0.92', 1.85', 3.70', 7.40', 11.20'$. Then, the following non-Gaussian statistics have been estimated:

\begin{enumerate}

 {\bf \item The skewness of the aperture mass map $\langle M_{ap}^3\rangle $} is the third-order moment of the aperture mass $M_{ap}(\theta)$ 
 and can be computed directly from shear maps filtered with different aperture mass.
The skewness is a measure of the asymmetry of the probability distribution function.
The probability distribution function will be more or less skewed positively depending on the abundance of dark matter haloes at the $\theta$ scale. 
The formalism exists to predict the skewness of the aperture mass map for a given cosmological model \citep[see e.g.][]{map:jarvis04,map:kilbinger05}.\\

 {\bf \item The kurtosis of the aperture mass map $\langle M_{ap}^4\rangle $} is the fourth-order moment of the aperture mass $M_{ap}(\theta)$
 and can be computed directly from the different aperture mass maps.
The kurtosis is a measure of the peakedness of the probability distribution function. 
The presence of dark matter haloes at a given $\theta$ scale will flatten the probability distribution function and widen its shoulders leading to a larger kurtosis.
The formalism exists to predict the kurtosis of the aperture mass map for a given cosmological model \citep{map:jarvis04,map:kilbinger05}.\\

 {\bf \item  The peak count of the aperture mass maps $P^{\mathcal{T}}_{M_{ap}}$}.
A peak is defined as connected pixels above a detection threshold $\mathcal{T}$. We consider all pixels that are connected via the sides or the corners of the pixel as one structure. It means, we are not discriminating between peaks due to massive halos and peaks due to projections of large-scale structures. The formalism exists to predict the peak counts in weak-lensing surveys, including the fraction of spurious detections caused by projections effects \citep[e.g.][]{peak:maturi10}.

\end{enumerate}

We have reviewed the state-of-the-art of the non-Gaussian statistics used to constrain cosmology.
Interesting analytical results relative to the shear three-point correlation function or the convergence bispectrum were also reported (e.g. \cite{three-point:ma00a,three-point:ma00b,three-point:scoccimarro01,three-point:cooray01}). However, when considering only the equilateral configuration of the bispectrum, it has been shown that the discrimination efficiency of the cosmological models is relatively poor \citep{stat:pires09b}. 
An analytical comparison has been performed in \cite{stat:berge10} (for an Euclid-like survey) between the full bispectrum and an optimal match-filter peak count and both approaches were found to provide similar results.
However, as the full bispectrum calculation has a much higher complexity than peak counting, and no public code exists to compute it (only an equilateral code is available (Pires et al. 2009a)), the bispectrum has not been considered in this study.

\subsection{Convergence statistics}
\label{sect_convstat}
\begin{figure}
\centerline{
\vbox{
\includegraphics[height=6.7cm,width=9.cm]{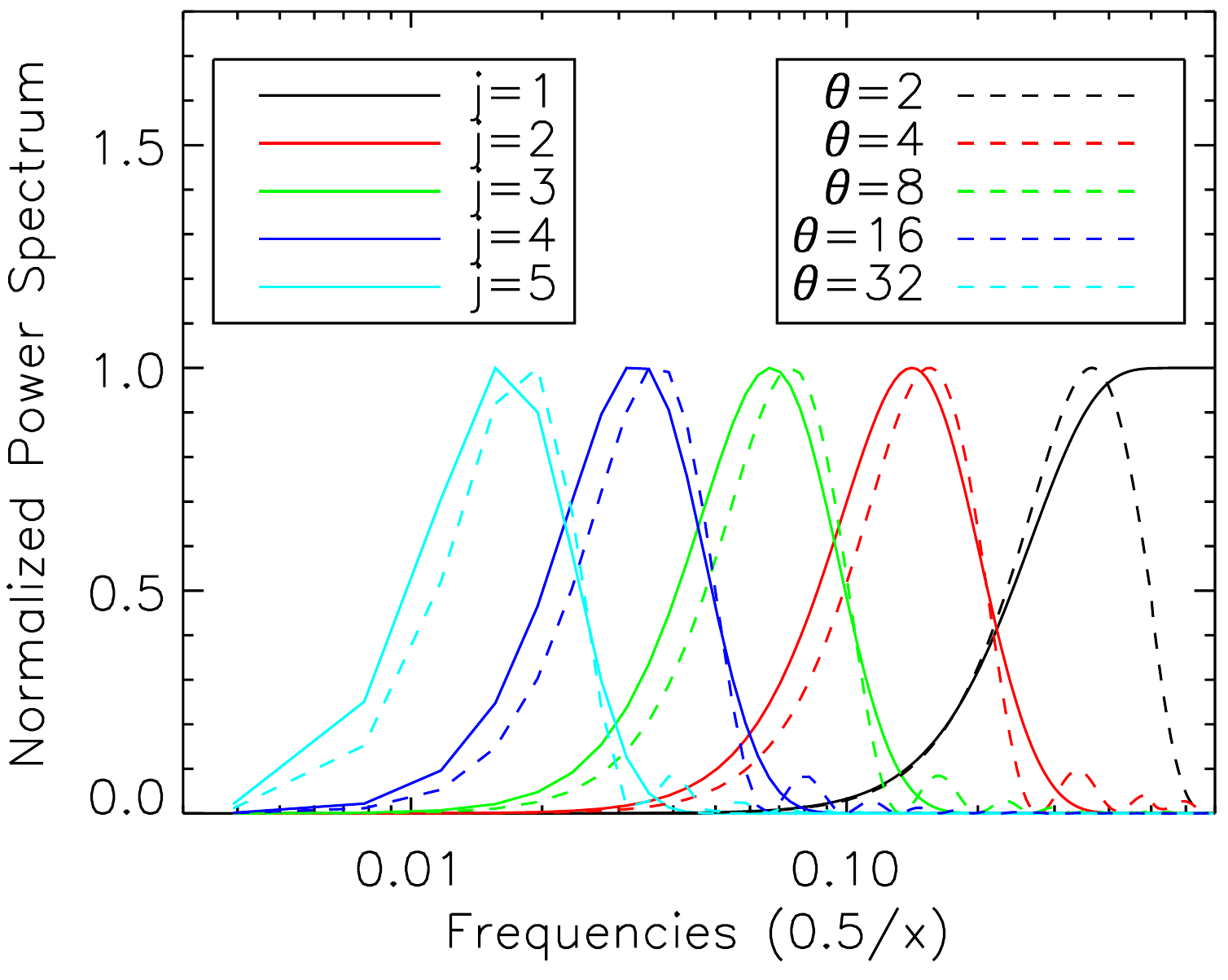}
}
}
\caption{Frequency response of the aperture mass filters for scales $\theta_i$= 0.46$\arcmin$, 0.92$\arcmin$, 1.85$\arcmin$, 3.70$\arcmin$, 7.40$\arcmin$ (solid lines) and frequency response of the wavelet filters at the same scales (dashed lines).}
\label{filters}
\end{figure}

In this section, we have used the noisy convergence maps described in \S \ref{sect_sim}.
The convergence has already been used in some studies \citep{stat:bernardeau97,stat:hamana04,stat:pires09a,stat:wang09,stat:berge10} to extract non-Gaussian information from higher-order statistics.
In this paper, we want to study the ability of higher-order shear statistics compared to higher-order convergence statistics to break the $\sigma_8$-$\Omega_m$ degeneracy. However, a fair comparison requires to compare the previous shear statistics with the convergence statistics at the same scale $\theta$ of the aperture mass.
We could have used the definition of the aperture mass expressed in terms of the convergence given by:
\begin{equation}
M_{ap}(\theta) = \int d^2 {\bf \vartheta} U_{\theta} (\vartheta) \kappa({\bf \vartheta}),
\label{map_conv}
\end{equation}
with:
\begin{equation}
U_{\theta}(\vartheta) = \frac{1}{2 \pi \theta^2} \left( 1- \frac{\vartheta^2}{2\theta^2} \right) \exp\left( -\frac{\vartheta^2}{2 \theta^2}\right).
\label{aperture_conv}
\end{equation}
However, we have preferred to use an undecimated isotropic wavelet transform: the "\`a trous" wavelet transform, which computes simultaneously $J$ aperture mass maps for dyadic scales. The { \bf "\`a trous" wavelet transform} decomposes a convergence map $\kappa$ (of size $N \times N$) as a superposition of the form:
\begin{equation}
\centering
\kappa(x, y) = c_J(x, y) + \sum_{j=1}^{J} w_j(x,y).
\label{eqn_wave}
\end{equation}
The algorithm outputs $J+1$ sub-band arrays of size $N \times N$
where $c_{J}$ is a coarse or smooth version of the original image $\kappa$
and $w_j$ represents the details of $\kappa$ at scale $2^{j}$ \citep[see][ for details]{starck:book98,starck:book02}.
\cite{map:leonard11} have shown that the wavelet bands $w_j$ are formally identical to aperture mass maps at scale $\theta = 2^j$ except the filter $U_{\theta}$ is replaced by the following wavelet function $\psi (x, y)$: 
\begin{equation}
\centering{
\frac{1}{4}\psi \left( \frac{x}{2},\frac{y}{2} \right)  =  \varphi(x,y) - \frac{1}{4}\varphi \left( \frac{x}{2}, \frac{y}{2} \right),}
\label{atrousw}
\end{equation}
where $\varphi(x,y)= \varphi(x)\varphi(y)$ and $\varphi(x)$ is a compact function (a B3-spline function) defined by:
\begin{eqnarray}
\varphi(x)=\frac{1}{12}(|x-2|^3-4|x-1|^3+6|x|^3-4|x+1|^3+|x+2|^3).\nonumber
\label{atrous5}
\end{eqnarray}
Fig. \ref{filters} displays in the Fourier domain in solid lines the aperture mass filters at scale $\theta_i = 2, 4, 8, 16, 32$ pixels (1 pixel = 0.23 arcmin) and in dashed lines, the corresponding wavelet filters at the same scale $2^j$ pixels with $j = 1, 2, 3, 4, 5$. 
To assess the response of aperture mass filters and wavelet filters, we have generated artificial shear data from a null convergence map with a single central delta function. The aperture mass algorithm has then been applied to the resulting shear maps and the wavelet transform has been computed from the convergence map. Then the response of these filters in Fourier space has been obtained by considering their power spectra.

These two filter banks are really close, except the last filters (from the left). 
The last wavelet filter is a high-pass filter whereas the last aperture mass filter is a band-pass filter. As a consequence, we can access to the finest scales of the image using the wavelet transform method which is not the case with the aperture mass method. Nevertheless, the generalized definition of the aperture mass statistics estimated from the shear n-point correlation functions \citep[see][]{map:schneider05,map:sembolini11} also gives access to these small scales because it is estimated on the shear catalogue directly. However, this has not been considered in this study because it is too intensive computationally. In the same way, there exists also some wavelet transforms that can work directly on the shear catalogue (see e.g. Deriaz et al., 2012). This has not been considered in this analysis because the resolution of the maps is very good and the noise is already dominant at this scale. 
Considering now the other filters, we see that the aperture mass filters have some unwanted oscillations that make the wavelet filters much more localized in Fourier space. These oscillations in Fourier space are due to the fact that the aperture mass $Q_{\theta}(\vartheta)$ defined in [\ref{aperture_shear}] is truncated for $\vartheta > \theta$ \citep[see][ for more details]{map:leonard11}.
Another point in favour of the wavelet transform is its time to compute. The wavelet transform complexity is $\varpropto \mathcal{O}(N^2 (j+1))$ compared to the aperture mass complexity, which is $\varpropto \mathcal{O}(N^2 \sum_i \theta_i^2)$. For one of our weak lensing simulations (of size $1024 \times 1024$), it is about $250$ times faster to compute the 5 considered scales with a wavelet transform than with the aperture mass definition.
The comparison between the aperture mass and the wavelet transform method has only been performed for scales : 0.92', 1.85', 3.70' and 7.40'. The scales in between have not been considered in this analysis because the dyadic scales are sufficient to characterize the variation of the discrimination as a function of the scale. Nevertheless, if we want to perform a more precise analysis, the scales in between this dyadic scaling can easily be obtained with the wavelet transform method by changing the initial pixel scale of the map.
In this study, for each simulated noisy convergence map $\kappa$ described in \S \ref{sect_sim}, we have estimated the wavelet transform from the previous definition [\ref{eqn_wave}] and then the following statistics have been estimated for each wavelet band $w_j$:

\begin{enumerate}
\item The skewness of the wavelet band $\langle w_j^3 \rangle$ that is computed directly from the different wavelet bands.
\item The kurtosis of the wavelet band $\langle w_j^4 \rangle$ that is computed directly from the different wavelet bands.
\item  The peak count of the wavelet band $P^{\mathcal{T}}_{w_j}$. A peak is defined as connected pixels above a detection threshold $\mathcal{T}$. 
\end{enumerate}

\subsection{Convergence statistics in denoised maps}
\label{sect_mrlensstat}

In this section, we want to derive higher-order statistics from denoised maps because the more the data are noisy and the more the probability distribution function looks like a Gaussian, the less the higher-order statistics will be useful. We expect, for example, the skewness and the kurtosis to tend to zero with an additive Gaussian noise and the clusters to be more difficult to extract. Therefore, to extract the non-Gaussian structures and reduce the impact of the noise in the analysis, we have used the MRLens denoising proposed by \cite{wlens:starck06}, which is a multiscale Bayesian denoising based on the sparse representation of the data. In \cite{wlens:starck06,wlens:teyssier09,stat:pires09a}, the authors have shown that this method outperforms several standard techniques to detect non-Gaussian structures such as Gaussian filtering, Wiener filtering and MEM filtering. \\

In this study, the MRLens denoising has been used to denoise each of the simulated convergence maps $\kappa$. The MRLens denoising software is available at the following address: "http://irfu.cea.fr/Ast/mrlens\_software.php". We have only applied this denoising to the noisy convergence maps because of the difficultly of applying denoising to the spin-2 shear field. Then, a wavelet transform has been applied to the denoised convergence map $\tilde \kappa$ in order to compute the denoised wavelet bands $\tilde w_j$ and estimate the following statistics:

\begin{enumerate}
\item The skewness of the denoised wavelet band $\langle \tilde w_j^3 \rangle $ that is computed directly from the denoised wavelet band $j$.
\item The kurtosis of the denoised wavelet band $\langle \tilde w_j^4 \rangle $ that is computed directly from the denoised wavelet band $j$.
\item The peak count of the denoised wavelet band $P^{\mathcal{T}}_{\tilde w_j}$. A peak is defined as connected pixels above a threshold $\epsilon$. Contrary to \S \ref{sect_shearstat} and \S \ref{sect_convstat} where the threshold $\mathcal{T}$ is used to extract peaks from the noise, the threshold $\epsilon$ (set to the rms value of the denoised wavelet band $\tilde{w}_j$) is only used to reject spurious detections in the denoised convergence maps.
\end{enumerate}

\section{Results}
\label{sect_res}
\subsection{The methodology}
\label{method}
In this study, we are interested in comparing the ability of the previous statistics to break the $\sigma_8$-$\Omega_m$ degeneracy, using the set of cosmological simulations described in \S \ref{sect_sim}. For this purpose, we have characterized, for each statistic, its abilty to discriminate between the 5 different cosmological models. As has been done in \cite{stat:pires09a}, we have computed a "discrimination efficiency" that expresses in percentage the ability of a statistic to discriminate between two cosmological models. For this purpose, a statistical tool called FDR (False Discovery Rate) introduced by \cite{fdr:benjamini95} has been used to set in an adaptive way the thresholds to classify between the different  cosmological models. Each threshold is estimated in such way that the rate of allowed false detections is inferior to a 0.05. The larger the discrimination efficiency is, the less the probability distributions of the statistic values for the different cosmological models overlap. The optimal statistic will be the one that maximizes the discrimination for all pairs of models. 
A mean discrimination efficiency can be estimated for each statistic by averaging the discrimination efficiency across all the pairs of models.

\subsection{Shear statistics results}
\label{sect_shearres}

Table \ref{disc_space_am} and Table \ref{disc_ground_am} show the mean discrimination efficiency for the shear statistics, described in \S \ref{sect_shearstat}, estimated for various aperture mass $M_{ap}(\theta_i)$ respectively for space-based and ground-based simulations.

\begin{table}
\begin{center}
\begin{tabular}{|c|c|c|c|c|c|}
\hline
$\theta_i$& $\langle M_{ap}^3\rangle $ &$\langle M_{ap}^4\rangle $& $P^{2 \sigma}_{M_{ap}}$  & $P^{3 \sigma}_{M_{ap}}$\\
\hline
0.46' & 04.60 \% & 02.30 \% & 39.70 \% & 54.95 \%  \\
\hline
0.69' & 23.85 \% &  02.25 \% &  67.05 \% & 55.10 \%  \\
\hline
0.92' & {\bf 33.40 \% }& 03.70 \% & 79.30 \% & 76.40 \%  \\
\hline
1.40' & 10.60 \% &  01.35 \% &  \fbox{ {\bf92.35 }} \% &  85.45 \%  \\
\hline
1.85' & 03.45 \% & 01.45 \% & 91.25 \% & 89.20 \%  \\
\hline
2.80' & 03.40 \% & 07.95 \% &  83.25 \% &  {\bf 90.00  \%}   \\
\hline
3.70' & 15.15 \% & 23.00 \% & 69.40 \% & 86.70 \%  \\
\hline
5.60' & 28.55 \% & {\bf 28.90 \%} &  02.45 \% & 70.55 \%  \\
\hline
7.40' & 26.95 \% &  24.30 \% & 4.90 \% & 60.50 \% \\
\hline
11.20' & 23.00 \% & 20.30  \% & 09.65 \% & 01.90 \%  \\
\hline
\end{tabular}
\end{center}
\caption{Mean discrimination efficiency (in percent) from noisy aperture mass maps (space-based simulations) for scales $\theta_i = 0.46', 0.92', 1.40', 1.85', 2.80', 3.70', 5.60', 7.40', 11.20'$.}
\label{disc_space_am}
\end{table}

\begin{table}
\begin{center}
\begin{tabular}{|c|c|c|c|c|c|}
\hline
$\theta_i$ & $\langle M_{ap}(\theta_i)^3\rangle $ &$\langle M_{ap}(\theta_i)^4\rangle $& $P^{2 \sigma}_{M_{ap}(\theta_i)}$  & $P^{3 \sigma}_{M_{ap}(\theta_i)}$\\
\hline
 0.46'  & 01.95 \% & 02.25 \% & 03.45 \% &06.50 \%  \\
\hline  
 0.69' & 01.75 \% &  01.05 \% & 12.50 \% & 05.60 \%  \\
\hline  
 0.92'  & 02.65 \% & 01.30 \% & 31.95 \% & 21.90 \%  \\
\hline  
 1.40' & 07.70 \% & 01.75 \% &  49.35 \% & 41.90 \%  \\
\hline
 1.85' & 04.05 \% & 02.15 \% &  {\bf  55.85 \%} & 48.85 \%  \\
\hline  
 2.80'  &  01.80 \% &  03.15 \% &  55.00 \% & \fbox{{\bf62.65 \%}}  \\
\hline
 3.70'  & 04.00 \% & 05.20 \% &  54.40 \% & 54.45 \% \\
\hline  
 5.60' & 09.70 \% & {\bf 10.90 \%} &  11.05 \% & 60.40 \%  \\
\hline
 7.40'  & {\bf 12.00 \% } & 10.25 \% & 14.15 \% & 51.90 \% \\
\hline  
11.20' & 09.05 \% &  08.90 \% & 04.30 \% & 09.60 \%  \\
\hline
\end{tabular}
\end{center}
\caption{Mean discrimination efficiency (in percent) from noisy aperture mass maps (ground-based simulations) for scales $\theta_i = 0.46', 0.92', 1.40', 1.85', 2.80', 3.70', 5.60', 7.40', 11.20'$.}
\label{disc_ground_am}
\end{table}

 \begin{table}
    \begin{tabular}{ c  c  c  c c}
      \hline
       &
      \multicolumn{2}{c @{}}{Space-based} &
      \multicolumn{2}{c}{Ground-based}
      \\
      \hline
       & 
       \multicolumn{1}{|c|}{Purity} &
       \multicolumn{1}{|c|}{Completeness} &
       \multicolumn{1}{|c|}{Purity} &
       \multicolumn{1}{|c|}{Completeness}
       \\
      \hline
      \hline
      $P^{2 \sigma}_{M_{ap}}$ & 16.07 \% & 61.27 \% &  10.77 \% & 35.60 \% \\       
      \hline
      $P^{3 \sigma}_{M_{ap}}$ & 47.63 \% & 35.11 \% & 33.15 \% & 11.56 \% \\
      \hline
       \end{tabular}
    \\
        \caption{Purity and completeness for the peak count for space-based (left) and ground based (right) aperture mass maps corresponding to realizations of the cosmological model with $\Omega_m = 0.3$ and $\sigma_8 = 0.9$). $P^{2 \sigma}_{M_{ap}}$ (respectively $P^{3 \sigma}_{M_{ap}}$) is defined for peaks above a $2\sigma$-threshold (respectively a $3\sigma$-threshold) on noisy aperture mass maps for scale $\theta=1.85'$.}
    \label{tablepcam}
  \end{table}

On first glance at these two tables, we can see that the results worsen with noise whatever the statistics. 
As expected, the skewness and the kurtosis are very poor at discriminating between different cosmological models in noisy aperture mass maps because the aperture mass map probability distribution function tends to a Gaussian distribution as the noise increases. This makes the skewness and the kurtosis of the aperture mass map tend to zero. The result with peak counting is significantly better because a basic denoising is applied by only selecting the peaks above a given threshold in the aperture mass map. However, the choice of this threshold is relatively important because the differences between a $2\sigma$ and a $3\sigma$ threshold are significant. 

Table \ref{tablepcam} shows the purity and the completeness for the peak count estimated on noisy aperture mass maps ($\theta=1.85'$) for space-based (left) and ground based (right). Completeness and purity are two important criteria to evaluate the performance of a peak detection method. Purity is defined as the ratio of true detections to the total number of peaks detected, and completeness is defined as the ratio of true detections to the total number of peaks in the simulation. 
The total number of peaks per scale $\theta$ is estimated from the simulated 2D convergence maps (without noise) with different aperture masses. It means, we are not discriminating between peaks due to massive halos and peaks due to projections of large scale-structures.
With a $2\sigma$-threshold, the completeness is maximal but the purity is poor because there is a large number of false detections due to shot noise, among the total number of detected peaks. Thus, the number of detected peaks is considerably overestimated especially at small scales for which the noise is important. Thus, the choice of the best threshold is a trade-off between purity and completeness.

Another important parameter is the scale $\theta$ of the aperture mass. 
The discrimination efficiency depends strongly on the scale that is considered. 
The best discrimination efficiency scale is displayed in bold, for each statistic, in Table \ref{disc_space_am} and Table \ref{disc_ground_am}.
We see that the best scale depends on the statistic that is used. 
This difference can be explained by the fact that the statistics are not sensitive in the same way to the different characteristics of the clusters.
Skewness and kurtosis are very sensitive to the density of the clusters, e.g. very dense clusters will skew significantly the probability distribution function whereas small clusters will have a small impact. In contrast, the peak count is mainly sensitive to the number of clusters regardless of their masses or their density. A massive cluster will be accounted in the same way as a small cluster if it is detected.

The best discrimination efficiency in noisy aperture mass maps (for space-based simulations) has been obtained with the peak count with a $2 \sigma$ threshold (92.35 \%) for a scale of $1.40 \arcmin$. This is definitely better than the best result obtained with the skewness (33.40 \%) as well as the best result obtained with the kurtosis (28.90 \%).

\subsection{Convergence statistics results}
\label{sect_convres}

\begin{table}
\begin{center}
\begin{tabular}{|c|c|c|c|c|c|}
\hline
Scale & $\langle w_j^3\rangle $ &$\langle w_j^4\rangle $& $P^{2 \sigma}_{w_j}$  & $P^{3 \sigma}_{w_j}$\\
\hline
Finest scales & 02.00 \% & 01.15 \% & 12.05\% & 00.70 \%  \\
\hline
0.92' & {\bf 37.95 \%} & 04.75 \% & 86.30 \% & 73.05 \%  \\
\hline
1.85' & 03.55 \% & 02.10 \% &  \fbox{{\bf 94.40  \%}} & {\bf  93.85 \% } \\
\hline
3.70' & 18.25 \% & 25.65 \% & 84.05 \% & 87.05 \%  \\
\hline
7.40' &   36.40 \% & {\bf 30.90} \%  & 24.60 \% & 66.35 \% \\
\hline
\end{tabular}
\end{center}
\caption{Mean discrimination efficiency (in percent) from noisy convergence wavelet maps  (space-based simulations) for the finest scales of the image given by the high-pass filter  (first column) and for scales $0.92', 1.85', 3.70',7.40'$ (other columns).}
\label{disc_space_wt}
\end{table}

\begin{table}
\begin{center}
\begin{tabular}{|c|c|c|c|c|c|}
\hline
Scale & $\langle w_j^3\rangle $ &$\langle w_j^4\rangle $& $P^{2 \sigma}_{w_j}$  & $P^{3 \sigma}_{w_j}$\\
\hline
Finest scales & 00.60 \% & 00.65 \% & 01.20 \% & 00.65 \%  \\
\hline
0.92' & 07.50 \% & 01.60 \% & 38.15 \% & 13.45 \%  \\
\hline
1.85' & 06.70 \% & 02.25 \% & \fbox{ {\bf 62.75 \%}}  & \bf{ 58.85 \% } \\
\hline
3.70' & 04.15 \% & 06.45 \% &  62.25 \%& 52.40 \%  \\
\hline
7.40' &  {\bf 14.55 \%} & {\bf17.65 \%} & 44.20 \% & 40.90 \% \\
\hline
\end{tabular}
\end{center}
\caption{Mean discrimination efficiency (in percent) from noisy convergence wavelet maps (ground-based simulations) for the finest scales of the image given by the high-pass filter  (first column) and for scales $0.92', 1.85', 3.70', 7.40'$ (other columns).}
\label{disc_ground_wt}
\end{table}

Table \ref{disc_space_wt} and Table \ref{disc_ground_wt} show the mean discrimination efficiency for the convergence statistics, described in \S \ref{sect_convstat}, estimated for different wavelet scales ($2^j$) respectively for space-based and ground-based simulations. 
As previously, the skewness and the kurtosis are very poor at discriminating between different cosmological models in noisy convergence maps because the skewness and the kurtosis tend to zero as noise is increased (see Fig. \ref{moments}).

A comparison with Table \ref{disc_space_am} and Table \ref{disc_ground_am} shows that the results obtained with the aperture mass maps  at the same scale are very similar. 
This is not a surprising result because it has been shown by \cite{map:leonard11} that applying aperture mass filters at dyadic scales in shear maps is comparable to performing a wavelet transform of the convergence map that can be derived from the shear maps by inversion in Fourier space. 
This also explains the similarity of the wavelet filters compared to the aperture mass filters at the same scales in Fig. \ref{filters}. However, the results are slightly improved with the wavelet transform for every statistic and every scale, which tends to show that the shape of the wavelet filters is more efficient to capture the non-Gaussian structures present in the weak lensing maps. This is possibly a consequence of the oscillations seen in the aperture mass filters (see Fig. \ref{filters}), which gives rise to a small leakage of the signal into higher frequencies.

{\bf Some other studies have been conducted to find an optimal filter for detecting dark matter haloes \citep{stat:maturi05,stat:pace07} and thus avoid the spurious peaks due to large-scale structure projections. However, these filters are less efficient because the projection effects that are normally a main source of uncertainty when probing the clusters, here serve as an additional source of cosmological information \citep[see][ for more details]{map:dietrich10,stat:wang09}.}

The best discrimination efficiency in noisy convergence maps (for space-based simulations) has been obtained with the peak count (94.40 \%), for a scale of $1.85 \arcmin$ and a $2 \sigma$-threshold. 

  \begin{table}
    \begin{tabular}{ c  c  c  c c}
      \hline
       &
      \multicolumn{2}{c @{}}{Space-based} &
      \multicolumn{2}{c}{Ground-based}
      \\
      \hline
       & 
       \multicolumn{1}{|c|}{Purity} &
       \multicolumn{1}{|c|}{Completeness} &
       \multicolumn{1}{|c|}{Purity} &
       \multicolumn{1}{|c|}{Completeness}
       \\
      \hline
      \hline
      $P^{2 \sigma}_{w_3}$ & 22.24 \% & 63.20 \% &  14.85 \% & 36.31 \% \\       
      \hline
      $P^{3 \sigma}_{w_3}$ & 56.66 \% & 38.76 \% & 42.02 \% & 13.11 \% \\
      \hline
      $P^{mrlens}_{w_3}$ & 84.30 \% & 49.55 \% & 75.37 \% & 25.92 \%\\
      \hline
    \end{tabular}
    \\
        \caption{Purity and completeness for the peak count for space-based (left) and ground-based (right) convergence maps corresponding to realizations of the cosmological model with $\Omega_m = 0.3$ and $\sigma_8 = 0.9$. $P^{2 \sigma}_{w_3}$ (respectively $P^{3 \sigma}_{w_3}$) is defined for peaks above a $2\sigma$-threshold (respectively a $3\sigma$-threshold) on noisy convergence maps at the third scale of a wavelet transform (1.85') and $P^{mrlens}_{w_3}$ is defined for peaks above a $\epsilon$-threshold on MRLens denoised convergence maps at the third scale of a wavelet transform (1.85').}
    \label{tablepcwt}
  \end{table}

Table \ref{tablepcwt} shows the purity and the completeness for the peak count at the third scale of a wavelet transform (1.85') for space-based (left) and ground-based (right) convergence maps. A comparison with Table \ref{tablepcam} shows that both purity and completeness are improved with the wavelet transform. As with the aperture mass statistic, the completeness is maximal with a $2\sigma$-threshold.

As previously, the constraints on cosmological models obtained with peak count (94.40 \%) are significantly better than the ones that can be reached with the  skewness (37.95 \%) and the kurtosis (30.90 \%).


\subsection{Denoised Convergence statistics results}
\label{sect_mrlensres}
In this section, we want to show that the convergence statistics can be improved significantly if denoising is applied to the convergence maps. As said previously, the convergence maps have been denoised using the MRLens denoising described in \cite{wlens:starck06}. Table \ref{disc_space_mrlens} and Table \ref{disc_ground_mrlens} show the mean discrimination efficiency for the denoised convergence statistics, described in \S \ref{sect_mrlensstat}, estimated for different wavelet scales ($2^j$) respectively for space-based and ground-based simulations.

\begin{table}
\begin{center}
\begin{tabular}{|c|c|c|c|c|c|}
\hline
Scale & $\langle \tilde w_j^3\rangle $ &$\langle \tilde w_j^4\rangle $& $P_{\tilde w_j}$ \\
\hline
Finest scales & 53.40 \% & 43.20 \% & 68.35 \% \\
\hline
0.92' & 47.90 \% & 41.15 \% & 92.45 \% \\
\hline
1.85' & 58.80 \% & 44.70 \% & \fbox{{\bf 96.75 \%}} \\
\hline
3.70' & { \bf 63.30 \%} & { \bf 48.05} \% & 90.40 \%  \\
\hline
7.40' & 54.90 \% & 40.45 \% & 63.45 \% \\
\hline
\end{tabular}
\end{center}
\caption{Mean discrimination efficiency (in percent) from MRLens denoised convergence wavelet maps (space-based simulations) for the finest scales of the image given by the high-pass filter  (first column) and for scales $0.92', 1.85', 3.70',7.40'$ (other columns).}
\label{disc_space_mrlens}
\end{table}

\begin{table}
\begin{center}
\begin{tabular}{|c|c|c|c|c|c|}
\hline
Scale & $\langle \tilde w_j^3\rangle $ &$\langle \tilde w_j^4\rangle $& $P_{\tilde w_j}$ \\
\hline
Finest scales & 42.15 \% & 30.05 \% & 38.20 \% \\
\hline
0.92' & 35.95 \% & 28.60 \% & 40.45 \%  \\
\hline
1.85' & 31.65 \% & 20.85 \% & 62.35 \% \\
\hline
3.70' & 41.80 \% & 29.95 \% & \fbox{{ \bf 72.65 \%}}  \\
\hline
7.40' & { \bf 44.75 \%} &{ \bf 32.25 \%} & 54.55 \% \\
\hline
\end{tabular}
\end{center}
\caption{Mean discrimination efficiency (in percent) from MRLens denoised convergence wavelet maps (ground-based simulations) for the finest scales of the image given by the high-pass filter  (first column) and for scales $0.92', 1.85', 3.70', 7.40'$ (other columns).}
\label{disc_ground_mrlens}
\end{table}

\begin{table*}
\begin{center}
\begin{tabular}{|c|c|c|c|c|c|c|}
\hline
& model 1 & model 2 & model 3  &  model 4 & model 5\\
\hline
model 1 & x & 85 \% & 100 \% & 100 \% & 100 \%  \\
\hline
model 2 & 89 \% & x & 92 \% & 100 \% & 100 \% \\
\hline
model 3 & 100 \% & 92 \% & x &  89 \% & 100 \% \\
\hline
model 4 & 100 \% & 100 \% & 92 \% & x & 98 \%  \\
\hline
model 5 & 100 \% & 100 \%  & 100 \% & 98 \%  & x\\
\hline
\end{tabular}
\end{center}
\caption{Discrimination efficiency (in percent) between the 5 cosmological models obtained with the peak count on denoised convergence maps at the third scale (1.85') of a wavelet transform (space-based simulations).}
\label{tablepcmrlens3}
\end{table*}
 
\begin{figure}
\centerline{
\vbox{
\includegraphics[height=7.cm,width=8.6cm]{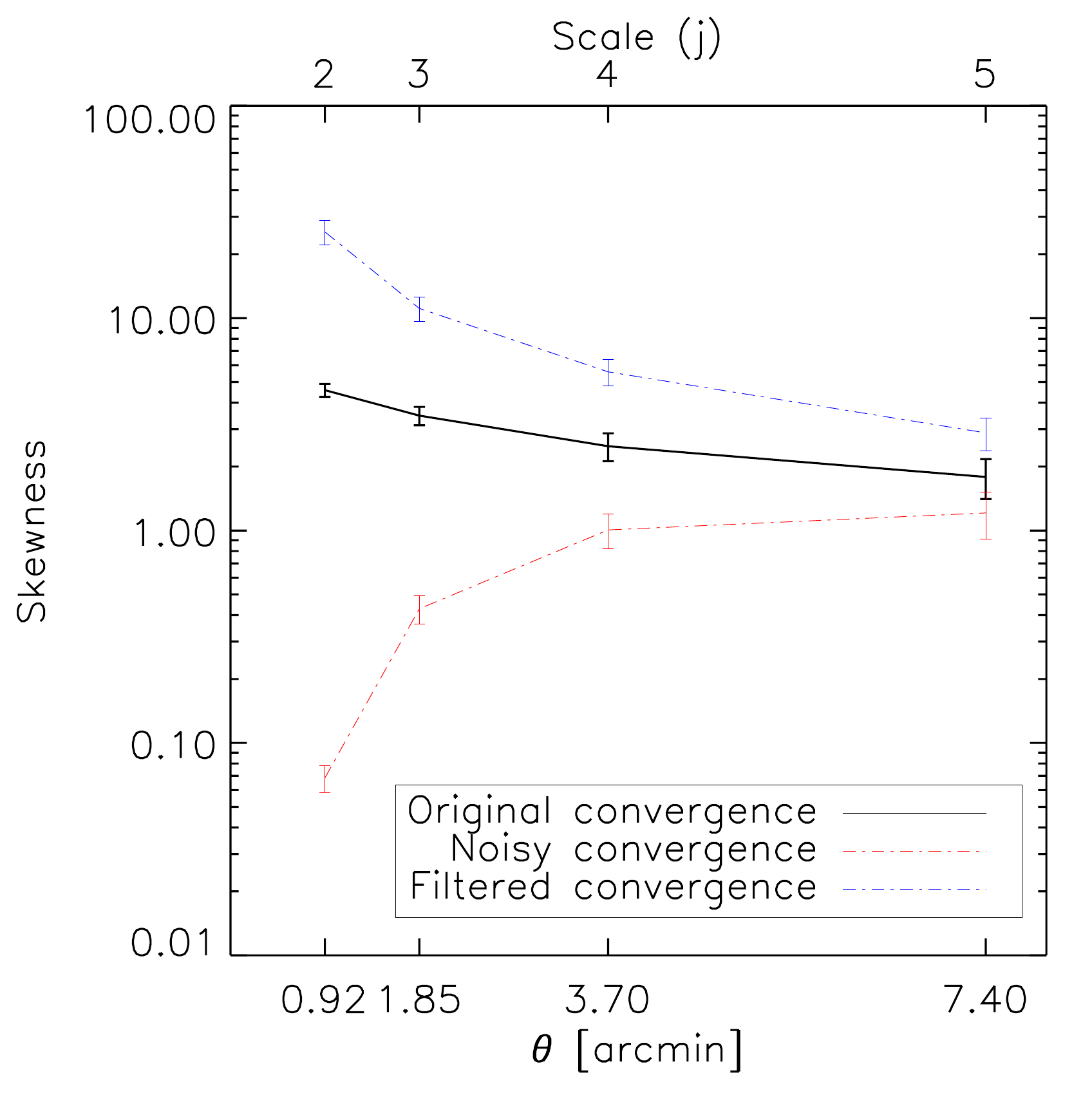}
\includegraphics[height=7.cm,width=8.5cm]{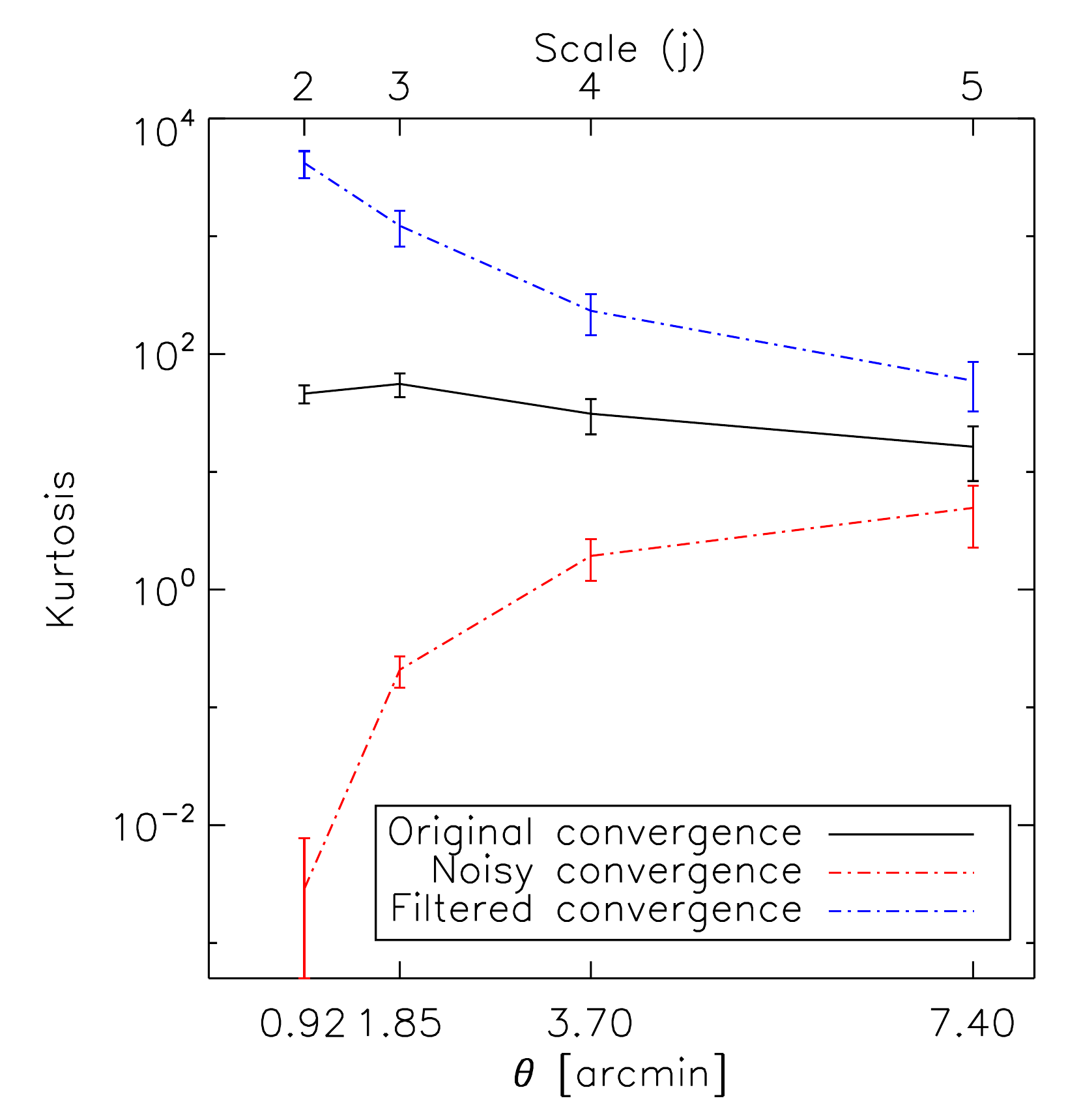}
}
}
\caption{Top: Mean skewness per scale for original convergence maps (black), space-based noisy convergence maps (red) and MRLens denoised convergence maps (blue). Bottom: Mean kurtosis per scale for original convergence maps (black), space-based noisy convergence maps (red) and MRLens denoised convergence maps (blue). The skewness and the kurtosis of the noisy convergence maps are considerably reduced especially at small scales for which the noise is important. In contrast, the skewness and the kurtosis are significantly overestimated on MRLens denoised convergence maps. These convergence maps correspond to realizations of the cosmological model with $\Omega_m = 0.3 $ and $\sigma_8 = 0.9$).}
\label{moments}
\end{figure}

As expected, the MRLens denoising improves considerably the discrimination efficiency of the skewness and kurtosis. This comes from its ability to reconstruct the non-Gaussian structures that dominate at small scales. However, the skewness and kurtosis values are significantly overestimated compared to original kurtosis, as shown in Fig. \ref{moments}, because the MRLens denoising is only efficient in recovering high peaks in the signal, which affects the tails of the probability distribution function. 

The MRLens denoising also improves the discrimination efficiency of the peak count at all scales, especially for ground-based simulations for which the noise is important. 

Table \ref{tablepcwt} shows the purity and the completeness for peak counting on MRLens denoised maps for space-based and ground-based simulations. 
In the MRLens denoising, we again have the usual trade-off between purity and completeness. A different threshold is selected for each wavelet band, and this is done in an adaptive way, conformed to a False Discovery Rate method \citep{wlens:starck06}, which provides a more robust discrimination. The completeness with MRLens denoising is slightly inferior to a $2\sigma$-threshold but its purity is maximal. 


The best discrimination efficiency in denoised convergence maps (for space-based simulations) has been obtained with the peak count (96.75 \%) still for a scale of $1.85 \arcmin$, in perfect agreement with the results of \cite{stat:pires09a}. Table \ref{tablepcmrlens3} shows the discrimination efficiency obtained with this statistic that enables to discriminate between the five cosmological models even for contiguous models for which the discrimination is challenging. The Table is not symmetric because the probability distributions of the statistics are not symmetric and to quantify the discrimination between two cosmological models, the FDR method sets two different thresholds in an adaptive way. This result can be compared with the result obtained in \cite{stat:pires09a} (Table 7) with another set of cosmological simulations. The results are very similar. The small discrepancies are only due to the limited size of the cosmological simulation sample.


\section{Conclusion}
\label{sect_cl}
The goal of this paper was to investigate how to best extract non-Gaussianity from weak lensing surveys to constrain the cosmological model.
For this purpose, we have been interested in showing that there is an extra information that can be derived from higher-order statistical analysis of the convergence maps beyond what can be obtained from higher-order statistical analysis of the shear maps. 
Therefore, we have compared the efficiency of several higher-order shear and convergence statistics to break the $\sigma_8-\Omega_m$ degeneracy, by comparing their ability to discriminate between 5 cosmological models along this degeneracy.

Most of the techniques used to estimate higher-order statistics from the spin-2 shear field are based on the aperture mass expressed in terms of the tangential component of the shear (see relation [\ref{map_shear}]). Analogous convergence statistics can be obtained by using the aperture mass defined from the convergence maps (see relation [\ref{map_conv}]). However, in accordance with \cite{map:leonard11}, we have preferred to use an alternative solution that computes simultaneously multiple aperture mass maps for dyadic scales: the "\`a trous" wavelet transform. 
In this study, we have observed that this method is 250 times faster than the aperture mass method to run on our simulations and that the wavelet filters are much more localized in Fourier space compared to aperture mass filters. It follows that the results obtained in wavelet convergence maps compared to the results in aperture mass maps are very similar but slightly improved in the wavelet case for every statistic and every scale. Therefore, contrary to a generally accepted idea, the noise properties in aperture mass maps are not better than in convergence maps, if the same scale is considered.

Contrary to another accepted idea, further important cosmological information can be extracted from noisy convergence maps if a denoising such as MRLens is used. This comes from its ability to reconstruct the non-Gaussian structures that are induced by the growth of structures. In this study, we have shown that the MRLens denoising improves considerably the discrimination efficiency of the skewness and kurtosis. It also improves the discrimination efficiency of the peak count especially for ground-based simulations ($n_g = 30$ gal/arcmin$^2$) for which the noise is important. 

For an Euclid-like survey, the density of galaxies is expected to be around $n_g = 40$ gal/arcmin$^2$ for the wide-field survey and around $n_g = 80$ gal/arcmin$^2$ for the deep-field survey. It is clear from this study, that the non-Gaussian statistical analysis should be performed in denoised convergence maps as described in \S \ref{sect_mrlensstat}.

Finally, the best non-Gaussian statistic to constrain cosmological model in combination with the power spectrum has been found to be the peak count per scale. 
And further cosmological information should be obtained by combining the constraints obtained with the peak count at different scale as shown by  \cite{stat:marian11}. This will be investigated in a future work.


\section*{Acknowledgments}
This work has been supported by the European Research Council grant SparseAstro (ERC-228261).

\bibliographystyle{astron}
\bibliography{biblio.bib}

\label{lastpage}

\end{document}